\documentclass[twocolumn,showpacs,preprintnumbers,amsmath,amssymb,12 pt.,double-spaced]{revtex4}

\usepackage{graphicx}
\usepackage{dcolumn}
\usepackage{bm}
\usepackage{color}

\begin{document}

\preprint{APS/123-QED}

\title{Kondo-like mass enhancement of Dirac fermion in iron pnictides Ba(Fe$_{1-x}$Mn$_x$As)$_2$}

\author{T. Urata$^1$}

\author{Y. Tanabe$^1$}
 \thanks{Corresponding author: youichi@sspns.phys.tohoku.ac.jp}

\author{K. K. Huynh$^2$}

\author{H. Oguro$^3$}

\author{K. Watanabe$^3$}

\author{S. Heguri$^2$}

\author{K. Tanigaki$^{1, 2}$}
\thanks{Corresponding author: tanigaki@sspns.phys.tohoku.ac.jp}

\affiliation{$^1$Department of Physics, Graduate School of Science, Tohoku University, Aoba, Aramaki, Aoba-ku, Sendai, 980-8578, Japan}

\affiliation{$^2$WPI-Advanced Institutes of Materials Research, Tohoku University, Aoba, Aramaki, Aoba-ku, Sendai, 980-8577, Japan}

\affiliation{$^3$High Field Laboratory for Superconducting Materials, Institute for Materials Research, Tohoku University, Sendai 980-8577, Japan}

\date{\today}

\begin{abstract}
The effect of Mn substitution, acting as a magnetic impurity for Fe, on the Dirac cone was investigated in Ba(Fe$_{1-x}$Mn$_x$As)$_2$.
Both magnetoresistance and Hall resistivity studies clearly indicate that the cyclotron effective mass ($m^{\ast}$) of the Dirac cone is anomalously enhanced at low temperatures by the impurity, although its evolution as a function of carrier number proceeds in a conventional manner at higher temperatures.
Kondo-like band renormalization induced by the magnetic impurity scattering is suggested as the most plausible explanation for this, and the anomalous mass enhancement of the Dirac fermions is discussed.

\end{abstract}

\pacs{74.70.Xa, 74.25.Dw, 72.15.Gd, 75.47.-m}

\maketitle

\section{Introduction}
The Dirac cone is a new paradigm in condensed matter physics, which has been identified so far in a variety of materials including graphene \cite{Castro}, topological insulators (TIs) \cite{Kane} and organic conductors \cite{Tajima}.
It has been noted that the conduction electrons in these Dirac cone system are characterized by especially long relaxation times.
More recently Dirac cone states have been found in the parent materials of iron pnictide superconductors \cite{Fukuyama, Ran, Morinari, Richard, Harison, Shimojima, Ming, Kim, Imai, YZhang}. 
Theoretical calculations for the Fe\,3$d$-multiband system predict different pseudospin vorticities originating from the uneven contribution of the Fe-3d orbitals to electron- and hole- pockets in the paramagnetic semimetal state \cite{Ran}.
At low temperatures ($T$s), a spin density wave (SDW) gap develops via the nesting between these electron- and hole- pockets, leading vorticity compensation between the pockets \cite{Ran, Morinari} and giving rise to a Dirac cone at the node of the SDW gap.
In contrast with other Dirac cone materials, density functional theory (DFT) calculations for these complex $d$-multiband iron pnictides indicate that the Dirac pockets are three-dimensional \cite{Yin}.
Accordingly, a systematic studies of the Dirac cone in this complex $d$-multiband material presents an opportunity to study a system of highly intriguing multibody interactions (Fig. 1 (a)).

\begin{figure}[t]
\includegraphics[width=0.95\linewidth]{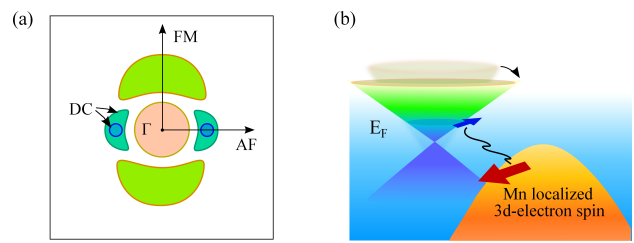}
\caption{Schematic images: (a) a ($k_x$, $k_y$) projection of the FS for Ba(FeAs)$_{2}$ in the spin-density-wave phase, as calculated by DFT \cite{Yin}; (b) the Kondo effect between Dirac fermions and 3d-Mn localized electron spins.}
\end{figure}

In Ba(FeAs)$_2$, linear magnetoresistance (LMR) was reported as a consequence of the quantum limit of the Dirac cone \cite{Khuong}.
This is typical for the Dirac cone as a result of its characteristic Landau level splitting, although it cannot be observed in a two-dimensional systems such as graphene as it is disrupted by quantum Hall oscillations \cite{Castro}.
The coexistence of a Dirac cone and high temperature superconductivity has been suggested a child compound of Ba(FeAs)$_2$, Ba(Fe$_{1-x}$Ru$_x$As)$_2$ \cite{Tanabe1}.
The Dirac cone in Ba(FeAs)$_2$ is robust for both nonmagnetic and magnetic impurities, since the Dirac cones in iron pnictide's complex $d$-multiband system have the same pseudospin chirality \cite{Morinari}.
Ba(Fe$_{1-x}$Ru$_x$As)$_2$ specifically showcases this resistance to nonmagnetic substitution impurities. Under Ru substitution, isoelectronic to the Fe-3$d$ state, the robustness of the Dirac cone leads to a suppression of backward scattering \cite{Tanabe2}.
Since the Dirac cones in iron pnictides originate from the pseudospin vorticities, they should also be resistant to magnetic impurities \cite{Morinari}, in contrast with topological insulators such as Bi$_2$Se$_3$ where the Dirac cone is disrupted by the breaking of time reversal symmetry \cite{Chen_TI}.
This unique resilience of the Dirac states in iron pnictides provides an attractive platform to study massless Dirac fermions, the Kondo effect \cite{Kondo}, and the Rudermann-Kittel-Kasuya-Yoshida (RKKY) interaction \cite{RKKY} in a correlated d-electron system, through the doping of magnetic impurities.

\begin{figure*}
\includegraphics[width=1.0\linewidth]{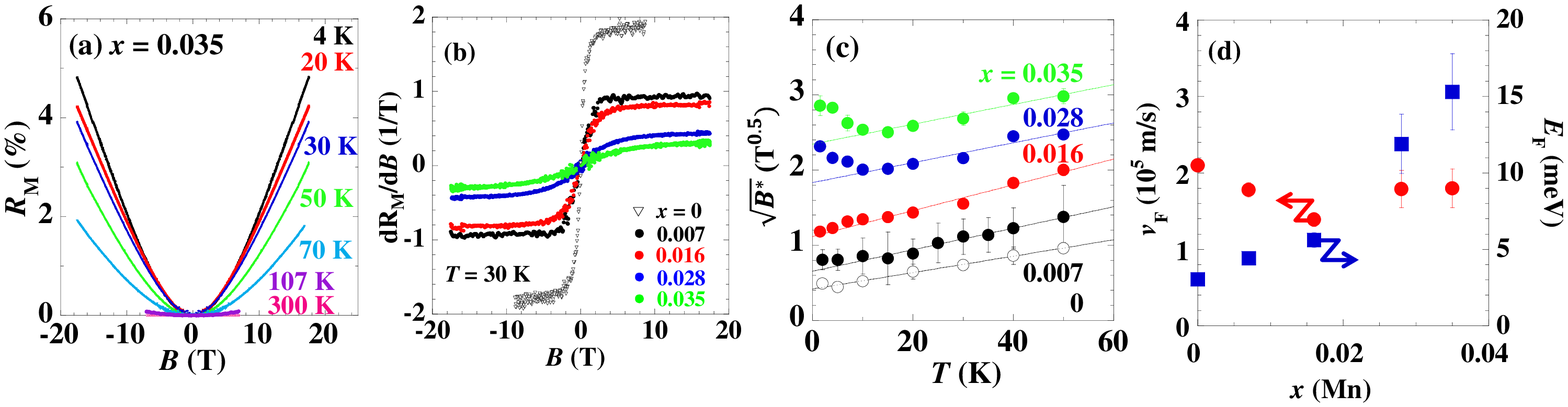}
\caption{(a) Magnetic field ($B$) dependence of magnetoresistance ($R_{\rm M}$) in Ba(Fe$_{1-x}$Mn$_x$As)$_2$ at various temperatures. (b) Derivative of $R_{\rm M}$ (d$R_{\rm M}$/d$B$) in Ba(Fe$_{1-x}$Mn$_x$As)$_2$ with $x$ = 0 - 0.035 at 30 K. (c) Temperature ($T$) dependence of the square root of $B^{\ast}$ in Ba(Fe$_{1-x}$Mn$_x$As)$_2$ with $x$ = 0 - 0.035. $B^{\ast}$ is defined as the onset field for the linear magnetoresistance. Solid lines indicate the fit versus eq.2 for $T$ $\geq$ 15 K. Error bars for $B^{\ast}$ were estimated from a least mean square fit of d$R_{\rm M}$/d$B$ as a function of B, where d$R_{\rm M}$/d$B$ develops linearly for low-$B$ and saturates for under high-$B$ (d) $x$ dependence of Fermi energy $E_{\rm F}$ and $v_{\rm F}$ for $x$ = 0 - 0.035. Error bars were estimated from a least mean squares fit of the $B^{\ast}$ versus $T$.}
\end{figure*}

Here, we report on the effect of Mn substitution on the Dirac cone in Ba(Fe$_{1-x}$Mn$_x$As)$_2$.
Nuclear magnetic resonance in Ba(Fe$_{1-x}$Mn$_x$As)$_2$ has shown that Mn substitution does not introduce any carrier doping but instead produces local magnetic moments \cite{Texier}.
Therefore, we regard Mn as a magnetic impurity for the Fe site in Ba(FeAs)$_2$. This compound has a complex $d$-multiband system, including both the Dirac cone and parabolic carrier pockets as shown in Fig. 1 (a).
To measure the electrical transport, single crystals of Ba(Fe$_{1-x}$Mn$_x$As)$_2$ were first grown, and then quenched from around 1323 K during a rapid cooling process. This introduces a lattice strain into the crystal structure. The samples obtained by such a procedure highlight the Dirac cone relative to the parabolic pockets, since the former is experimentally less sensitive to lattice distortion, while in the latter the relaxation times of the electrons are greatly decreased \cite{Tanabe2}.
We will show that although LMR is observed for Ba(Fe$_{1-x}$Mn$_x$As)$_2$, the temperature dependence of the onset magnetic field for LMR ($B^*$) departs significantly from the trend typically expected due to the Landau Level (LL) splitting of Dirac cones at low temperatures \cite{Tanabe1}.
At 2 K both the inverse of the carrier mobility (1/$\mu$) and the cyclotron effective mass ($m^{\ast}$) of the Dirac cone no longer obey the square root of the carrier number, instead exhibiting an unusual enhancement of $m^{\ast}$.
This anomalous LMR behavior in Ba(Fe$_{1-x}$Mn$_x$As)$_2$ will be described in detail, and discussed in terms of the Kondo effect between Dirac fermions and the local magnetic spins of the Mn impurities.

\section{Experiment}
Single crystals of Ba(Fe$_{1-x}$Mn$_x$As)$_2$ were grown by the flux method using FeAs \cite{Canfield}.
The quality of the single crystals was gauged by synchrotron X-ray diffraction, using the BL02B2 beam line at SPring-8.
Electric resistivity ($\rho$)  was also measured to confirm the quality of the samples.
The Mn concentration $x$ for Ba(Fe$_{1-x}$Mn$_x$As)$_2$ was determined based on the SDW transition temperature ($T_{\rm N}$), which has been measured previously for various values of $x$ in Ba(Fe$_{1-x}$Mn$_x$As)$_2$ \cite{Canfield} (Table I).
Further details of the characterization process for Ba(Fe$_{1-x}$Mn$_x$As)$_2$ are described in the appendix.
The dependence of both in-plane transverse magnetoresistance ($R_{\rm M}$) and Hall resistivity ($\rho_{yx}$) on $B$ were measured using a four-probe method for $B$ $\leq$ 17.5 T at various temperatures between 2 and 300 K.

\section{Results}
\subsection{Magnetoresistance and Hall resistivity}

Figure 2 (a) shows the $B$ dependence of $R_{\rm M}$ for Ba(Fe$_{1-x}$Mn$_x$As)$_2$ with $x$ = 0.035 at various temperatures.
Below the structural and magnetic transition temperature ($T_{\rm N}$ = 107 K), $R_{\rm M}$ varies markedly against $B$, in contrast to the relatively flat response observed at 300 K.
To examine the gradient of $R_{\rm M}$ versus $B$ in detail, the $B$ dependence of the derivative of $R_{\rm M}$ (d$R_{\rm M}$/d$B$) at 30 K for $x$ = 0 - 0.035 was measured as shown in Fig.2(b).

Although $R_{\rm M}$ decreases with increasing $x$, $R_{\rm M}$/d$B$ develops linearly against $B$, saturating above the crossover magnetic field ($B^*$).
Furthermore, we note that $B^*$ increases with increasing $x$.

\begin{figure*}
\includegraphics[width=1.0\linewidth]{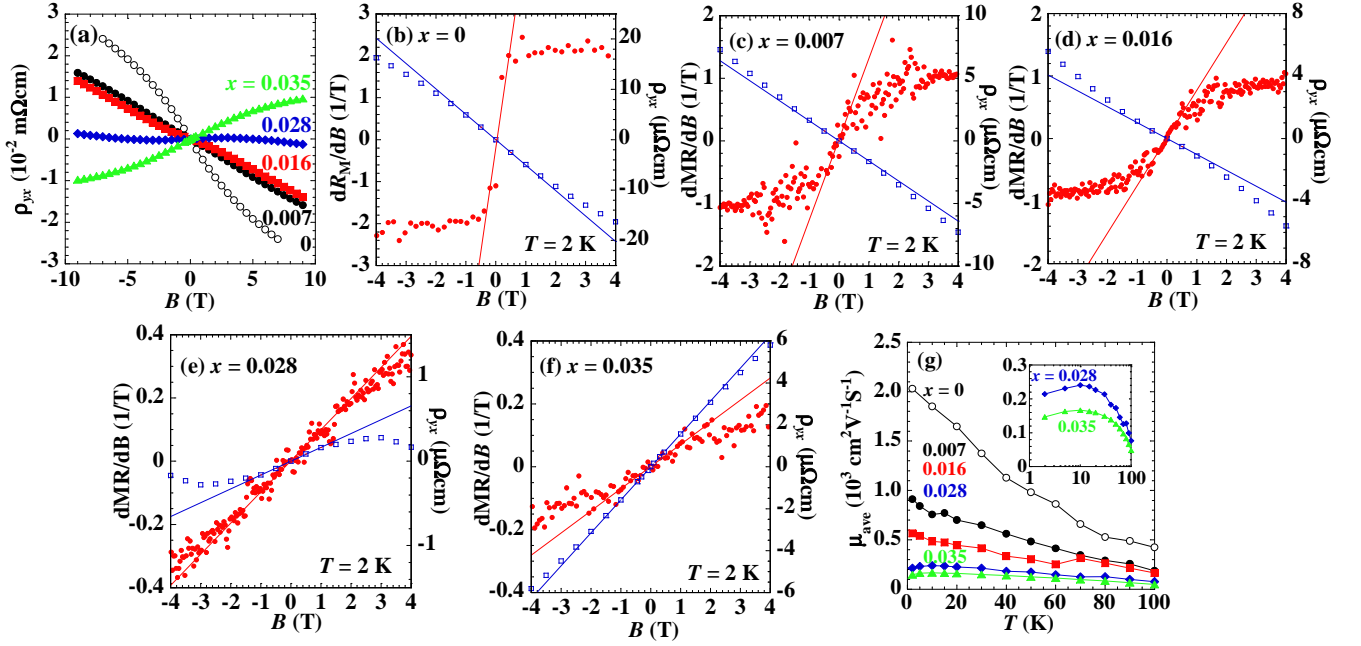}
\caption{(a) Magnetic field ($B$) dependence of Hall resistivity ($\rho$$_{yx}$) for Ba(Fe$_{1-x}$Mn$_x$As)$_2$ at 2 K for $x$ = 0 - 0.035. (b) - ((f) Magnetic field ($B$) dependence of the 1st derivative of magnetoresistance (d$R_{\rm M}$/d$B$), and the Hall resistivity ($\rho$$_{yx}$) from -4 to 4 T. (g) Temperature ($T$) dependence of averaged mobility ($\mu$$_{ave}$ = ($\mu_e$ + $\mu_h$)/2) for Ba(Fe$_{1-x}$Mn$_x$As)$_2$ with $x$ = 0 - 0.035. The inset in (g) shows a magnified plot for $x$ = 0.028 and 0.035.}
\end{figure*}

Figure 3 (a) shows the $B$ dependence of $\rho$$_{yx}$ for $x$ = 0 - 0.035 at 2 K.
While non-linear $\rho$$_{yx}$ behavior can be observed in the high $B$ regime, $\rho$$_{yx}$ develops linearly against $B$ in the low $B$ regime.
For larger values of $x$, the gradient of $\rho$$_{yx}$ under low $B$ increases, positive for the $x$ = 0.035 sample.
This can be understood easily as an increase in hole mobility with increasing $x$ in the low $B$ regime, as expected within the framework of a semiclassical approximation \cite{AM}.
In Figure 3 (b) - (f) the $B$ dependence of the 1st derivative d$R_{\rm M}$ (d$R_{\rm M}$/d$B$), and $\rho$$_{yx}$ are plotted in order to illustrate the low-$B$ behavior.
Both d$R_{\rm M}$/d$B$ and $\rho$$_{yx}$ are linear within the same $B$ regime, again consistent with the semiclassical approximation for a multi carrier system in the low-$B$ limit \cite{AM}.
For $x$ = 0.028, $\rho$$_{yx}$ deviates from its linear behavior at a lower value of $B$ than d$R_{\rm M}$/d$B$.
This may originate from the strong compensation between holes and electrons at low field strength, resulting in the sign change in $\rho$$_{yx}$.
It is noted that such compensation between electrons and holes does not induce a sign change in $R_{\rm M}$, but does lead to a deviation from the ideal quadratic $R_{\rm M}$ relationship.
While both d$R_{\rm M}$/d$B$ and $\rho$$_{yx}$ are affected by scattering in the low-$B$ regime, d$R_{\rm M}$/d$B$ is less sensitive to it, resulting in a wider $B$-linear regime in d$R_{\rm M}$/d$B$ than in $\rho$$_{yx}$.

\subsection{Analysis of the linear magnetoresistance; Quantum limit of Dirac cone}

Linear magnetoresistance was observed above $B^{\ast}$ in the SDW ordering state of Ba(Fe$_{1-x}$Mn$_x$As)$_2$ for $x$ = 0 - 0.035.
Generally, quantized cyclotron motion of electrons generates Landau level (LL) splitting, with all carriers occupying the 0th LL in the high field limit.
Under such a quantum limit, LMR is theoretically predicted for Dirac cone linear dispersion \cite{Abrikosov}, and has been experimentally confirmed in Ba(FeAs)$_2$ \cite{Khuong}.

The formulas for the energy splitting $\Delta_{\rm LL}$, the cyclotron mass $m^*$, and $B^*$ are shown as below \cite{Khuong, Castro, Abrikosov}:
\begin{eqnarray}
 \Delta_{\rm LL} &=& \displaystyle{\pm v_{\rm F}\sqrt{2{\rm e} {\rm \hbar} B}} \label{1},\\
 B^* &=& \displaystyle{(1/2{\rm e} {\rm \hbar} v_{\rm F}^2)(E_{\rm F} + k_{\rm B}T)^2} \label{2},\\
 m^{\ast} &=& \displaystyle{E_{\rm F}/v_{\rm F}^2 = \sqrt{\pi n_{\rm D}}/v_{\rm F}} \label{3}.
\end{eqnarray}
The energy splitting between the 0th and the 1st LL's of the Dirac cone is described by Eq.1 \cite{Abrikosov}, where $E_{\rm F}$ and $v_{\rm F}$ are the Fermi energy and the Fermi velocity of the Dirac cone.
At the quantum limit, the splitting between 0th and 1st LL's ($\Delta_{\rm LL}$) becomes larger than $E_{\rm F}$.
The crossover magnetic field $B^{\ast}$, where the magnetoresistance changes from semiclassical to quantum behavior, is described in Eq. 2 \cite{Khuong}.
This characteristic condition for LMR gives direct evidence for the quantum limit of the Dirac cone.
Figure 2(c) shows the $T$ dependence of $\sqrt{B^{\ast}}$ for $x$ = 0 - 0.035.
The curves for $x$ = 0 - 0.035 are fitted by Eq. 2 at temperatures above 15 K and were shown as solid lines in Fig. 2(c).
For $x$ = 0 - 0.016, $\sqrt{B^{\ast}}$ develops linearly with increasing $T$, and shows good agreement with Eq.\,2.
This is consistent with the picture of the quantum limit of the Dirac cone.
Although $\sqrt{B^{\ast}}$ agrees with Eq. 2  for $x$ = 0.028 and 0.035 above 15 K, $\sqrt{B^{\ast}}$ exhibits a sudden and large increase below 15 K, deviating from Eq. 2 at low temperatures.
Since the gradient of $\sqrt{B^{\ast}}$ was determined by $E_{\rm F}$ and $v_{\rm F}$, this low $T$ deviation is tied to a change in $E_{\rm F}$ and $v_{\rm F}$ within the framework of Eq. 2.
Figure 2(d) shows the estimated values for $E_{\rm F}$ and $v_{\rm F}$ derived from the fitting of the $T$ dependent $\sqrt{B^{\ast}}$ above 15 K by Eq.\,2.
$E_{\rm F}$ continuously increases with increasing $x$, while $v_{\rm F}$ remains roughly constant in the range between 1.4 - 2.0 $\times$ 10$^5$ m/s.

\begin{figure*}
\includegraphics[width=1.0\linewidth]{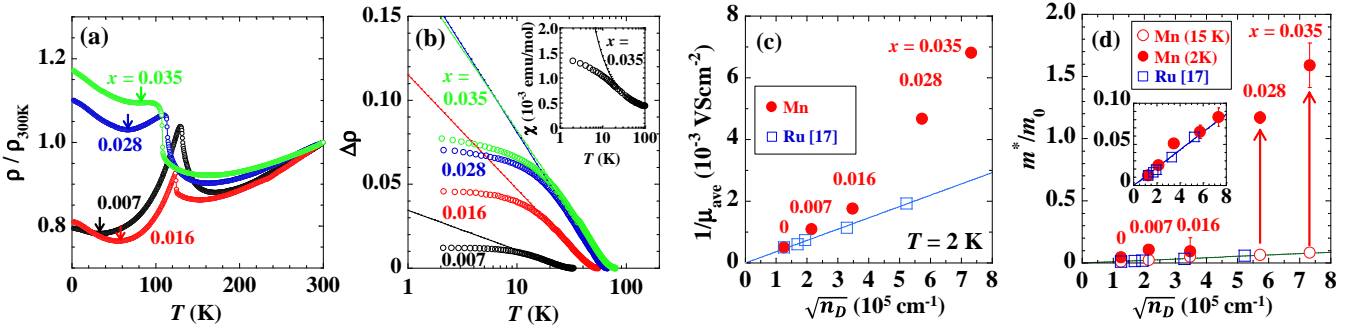}
\caption{(a) $T$ dependence of electrical resistivity ($\rho$) divided by $\rho$ at 300 K ($\rho$/$\rho_{300 K}$) for $x$ = 0.007 - 0.035. Arrows indicate the local minima of $\rho$ ($\rho$$_{\rm m}$).
(b) $T$ dependence of $\Delta$$\rho$ (= ($\rho$ - $\rho$$_{\rm m}$)/$\rho$$_{300 K}$). The inset of (b) shows the $T$ dependence of the magnetic susceptibility for $x$ = 0.035. A fit to the Curie-law ($\chi$ = $\chi$$_0$ + C/$T$; C : Curie constant) is indicated by the solid line.
(c) Inverse of $\mu_{\rm ave}$ (1/$\mu_{\rm ave}$) as a functon of the square root of carrier number for the Dirac cone ($n_{\rm D}$$^{0.5}$) of Ba(Fe$_{1-x}$Mn$_x$As)$_2$. For comparison, the dependence of Ba(Fe$_{1-x}$Ru$_x$As)$_2$ is also shown \cite{Tanabe2}. Solid lines are the results of a linear fit for Ba(Fe$_{1-x}$Ru$_x$As)$_2$. The error bars of 1/$\mu_{\rm ave}$ are calculated from the least squares error of the semiclassical tranport parameters. The estimated error of less than 1 $\%$ is much less than the difference of 1/$\mu_{\rm ave}$ between the Mn and Ru substituted samples.
(d) The cyclotron effective mass as a function of $n_{\rm D}$$^{0.5}$ at 2 K ($m^{\ast}$$_{2 K}$) for Ba(Fe$_{1-x}$Mn$_x$As)$_2$ (red open circle). The dependence of $m^{\ast}$ estimated from the parameters as shown in Fig.\,2(d) is also plotted (red closed circle). The $m^{\ast}$ values for Ba(Fe$_{1-x}$Ru$_x$As)$_2$ are also shown \cite{Tanabe2} (closed square). A linear fit for Ba(Fe$_{1-x}$Ru$_x$As)$_2$ is indicated by the solid line. Error bars for $m^{\ast}$ are calculated based on the error of $E_{\rm F}$ and $v_{\rm F}$.}
\end{figure*}

\subsection{Estimation of carrier mobilities; two-carrier-type semiclassical model}
It was found that LMR for $x$ = 0 - 0.035 is consistent with the quantum limit of the Dirac cone.
Mn substitution was found to increase the carrier number of the Dirac cone, but $v_{\rm F}$  remained nearly constant.
It therefore appears that although the Mn substitution changed the energy level of the Dirac neutral point, it did not change the total carrier number due to the compensation between the parabolic bands and the Dirac cones.
We conclude from this that the linear dispersion of Dirac cone is not disturbed by Mn substitution.
On the other hand, $\sqrt{B^{\ast}}$ showed an immediate increase below 15 K for $x$ = 0.028 and 0.035.
This suggests that $E_{\rm F}$ increases and $v_{\rm F}$ decreases below 15 K.
For the Dirac cone, the cyclotron effective mass ($m^{\ast}$) can be calculated using Eq. 3 \cite{Castro}.
Because $m^{\ast}$ is described by these two parameters, the increase in $B^{\ast}$ may indicate a decrease in mobility ($\mu$) as described by Eqs.(2) and (3).
In order to confirm the above scenario, we estimated the mobility based on a two-carrier-type semiclassical approximation in the low $B$ limit \cite{AM}.
At this limit, the zero-field resistivity ($\rho$), $R_{\rm M}$ and $\rho$$_{yx}$ are given by:
\begin{eqnarray}
 \rho &=& \frac{1}{e(n_e\mu_e + n_h\mu_h)} \label{4},\\
 R_{\rm M} &=& \frac{\rho(B) - \rho(0)}{\rho(0)} = \frac{n_en_h\mu_e\mu_h(\mu_e + \mu_h)^2B^2}{(n_e\mu_e + n_h\mu_h)^2} \label{5},\\
 \rho_{yx} &=& \frac{(-n_e\mu_e^2 + n_h\mu_h^2)B}{e(n_e\mu_e + n_h\mu_h)^2}. \label{6}
\end{eqnarray}
where $n_{\rm e}$ and $n_{\rm h}$ are the carrier number of holes and electrons, and $\mu$$_e$ and $\mu$$_h$ are their mobilities.
In order to estimate these four transport parameters using the above three equations, one additional constraint is necessary.
It was reported by angle-resolved photo emission spectroscopy (ARPES) that the Luttinger volumes of both electron and hole Fermi surfaces are comparable to each other in Ba(FeAs)$_2$ \cite{Shen} and that the Mn substitution does not result in additional carriers \cite{Texier}.
Therefore, we can approximate the system as a semimetal, that is, $n_{e}$ = $n_{h}$.
We then systematically solve $\rho$, $R_{\rm M}$ and $\rho$$_{yx}$ using Eqs. (1)\,-\,(3), where $R_{\rm M}$ and $\rho$$_{yx}$ are quadratic and linear against $B$, respectively.

For Ba(FeAs)$_2$, it has been theorized that the Fermi surface under the SDW ordered state is composed of four pockets \cite{Yin}.
If this is the case, the transport parameters in each pocket of the four carrier system can be merged into $n_{e}$, $n_{h}$, $\mu_{e}$, and $\mu_{h}$ within a two carrier model \cite{Tanabe2}.
We thus use the averaged mobility ($\mu_{\rm ave}$ = ($\mu$$_{\rm e}$ + $\mu$$_{\rm h}$)/2) to evaluate the dependence of mobility on the carrier number of the Dirac cones.

Figure 3(g) shows the carrier mobility $\mu_{\rm ave}$ estimated in this way.
Below $T^{\ast}$, $\mu_{\rm ave}$ continuously increased with decreasing $T$ for $x$ = 0 - 0.016.
It is noted that $\mu_{\rm ave}$ for $x$ = 0.028 and 0.035 saturate around 10 K, before decreasing at lower $T$s.

\section{Discussion}
The decrease in $\mu_{\rm ave}$ below 10\,K for $x$ = 0.028 and 0.035 is qualitatively consistent with the $T$ dependence of $B^{\ast}$ described by the Abrikosov model \cite{Abrikosov}.
It was reported that artificial defects act as local magnetic impurities in graphene, leading to the Kondo behavior \cite{Chen}.
Iron pnictides may exhibit a similar behavior, since magnetic impurity scattering does not break the Dirac cone states due to the pseudospin chiral symmetry \cite{Morinari}.
As such, one might propose that the Kondo effect emerges between the Dirac fermions and the localized 3$d$-electron spins arising from the Mn impurities.
In fact, previous studies have recently reported the Kondo effect in iron pnictides, supporting the realisity of the present assumption \cite{Wang, Tarantini}.
In order to search for the behaviors which characterize the Kondo effect, we studied the $T$ dependence of $\rho$ (normalized as $\rho$/$\rho$$_{300 K}$) for $x$ = 0.007 - 0.035, as shown in Fig.\,4(a).
Under the SDW states where the Dirac cone forms, the normalized resistivity ($\rho$/$\rho$$_{300 K}$) shows a local minima ($\rho$$_{\rm m}$) before increasing again when $T$ is lowered further \cite{Canfield}.
Similar data was published by Thaler et al.\cite{Canfield}, but their discussion focused mainly on the differences between Mn- and Co- substitutions from the viewpoint of supercoductivity.
In order to study the behavior of $\rho$/$\rho$$_{300 K}$ at low $T$, we plotted $\Delta$$\rho$ (($\rho$ - $\rho$$_{\rm m}$)/$\rho$$_{300 K}$) below temperature of the local minima $\rho$$_{\rm m}$ in Fig.\,4 (b).
Intriguingly, $\Delta$$\rho$ increases logarithmically as T decreases, saturating in the low $T$ limit.
This behavior is consistent with the picture of the Kondo effect for a system consisting of dilute magnetic impurities and conduction electrons.
Moreover, the $T$ dependence of the magnetic susceptibility ($\chi$) for $x$ = 0.035  deviates from the Curie-law below 30 K as shown in the inset of Fig.\,4 (b) and shoes saturation at low-$T$. 
This is consistent with quasiparticle mass enhancement via the Kondo effect.

\begin{figure*}
\includegraphics[width=1.0\linewidth]{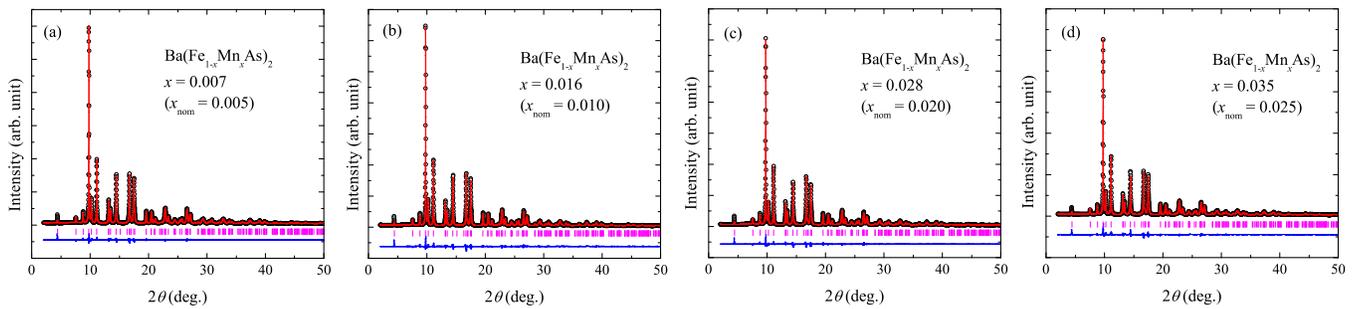}
\caption{(color online) Powder x-ray diffraction patterns for Ba(Fe$_{1-x}$Mn$_x$As)$_2$ with $x$ = (a) 0.007, (b) 0.016, (c) 0.025, and (d) 0.035. The open black circles indicate the the observed intensities. The the red lines are their Rietveld fits. Pink bars lines correspond to the Bragg-peak positions of Ba(FeAs)$_2$. The blue curve shows the calculated differences between the Rietveld fitting and the observed intensities.}
\end{figure*}

In a Kondo system with conduction electrons and localized electrons ($S$ = 1/2), the Kondo singlet forms as the ground state, leading to an increase in $m^{\ast}$ \cite{Yoshida}.
Since the present $B^{\ast}$ was directly tied to the quantum behavior of the Dirac cone, and $\mu_{\rm ave}$ is likewise influenced by the Dirac cone in the rapidly quenched Ba(FeAs)$_2$ series \cite{Tanabe2}, the present observations, showing that $B^{\ast}$ increases and $\mu_{\rm ave}$ decreases, can be associated with a Kondo-like mass enhancement of the Dirac fermions in Ba(Fe$_{1-x}$Mn$_x$As)$_2$ under the Abrikosov model \cite{Abrikosov}(Fig. 1(b)).
Recently, in order to explain LMR in iron pnictides, Koshelev theoretically investigated the realization of LMR under the nodal-SDW state \cite{Koshelev}.
In his theory, Koshelev notes out that $B^{\ast}$ is proportional to both 
the SDW gap and the inverse of the quasiparticle scattering rate, resulting in a nonmonotonic $T$ dependence of $B^{\ast}$.
However, $\sqrt{B^{\ast}}$$\propto$$T$ has been experimentally confirmed in Ba(FeAs)$_2$ systems within a wide $T$ range \cite{Khuong, Tanabe1}, which is likewise consistent with LMR in the Dirac cone \cite{Abrikosov}.
In the present work we also observe an increase in $B^{\ast}$ in the $\Delta$$\rho$ saturation region below 10 K for $x$ = 0.028 and 0.035 samples, in disagreement with the model proposed by Koshelev.
It should be noted that the Kondo temperature may be high ($\sim$ 20 - 30 K) considering the $T$ dependence of $\Delta$$\rho$ shown in Fig.4(b), while the Kondo screening can be efficiently suppressed by the Fe 3d electron ordering in iron pnictides \cite{Dai}.
Since an anomolously high Kondo temperature ($\sim$ 100 K) has also been reported in graphene \cite{Chen}, the nature of the Kondo effect between the Dirac fermions and the localized 3d electrons of Mn provides an interesting point for scientific debate.

In order to evaluate the enhancement of $m^{\ast}$ for the Dirac fermions in greater detail, 1/$\mu_{\rm ave}$ was plotted versus the square root of the Dirac cone carrier number ($\sqrt{n_{\rm D}}$), as shown in Fig.\,4(c).
Given that $n_{\rm D}$ could not vary below $T^{\ast}$ it was estimated using Eq.3, from the values of $E_{\rm F}$ and $v_{\rm F}$ shown in Fig.\,2(d).
For comparison, the dependence of $\sqrt{n_{\rm D}}$ on 1/$\mu_{\rm ave}$ is also shown for Ba(Fe$_{1-x}$Ru$_x$As)$_2$.
We can easily recognize a remarkable contrast between the linear evolution of 1/$\mu_{\rm ave}$ observed for Ba(Fe$_{1-x}$Ru$_x$As)$_2$ and the positive deviation seen in Ba(Fe$_{1-x}$Mn$_x$As)$_2$.
The former has been consistently explained in terms of the dependence of $m^{\ast}$ on $n_{\rm D}$ in the Dirac cone \cite{Castro, Tanabe2}.
Given the invariant $n_{\rm D}$ of the Dirac cone for Ba(Fe$_{1-x}$Mn$_x$As)$_2$, $m^{\ast}$ at the lowest temperature of 2 K was evaluated using Eqs.2 and 3.
Figure 4(d) shows $m^{\ast}$ normalized by the free-electron mass ($m_0$) as a function of $\sqrt{n_{\rm D}}$ at 2 K, along with $m^{\ast}$/$m_0$ at 15 K estimated from values shown in Fig.\,2(d).
The dependence of $m^{\ast}$ /$m_0$ on $\sqrt{n_{\rm D}}$ for Ba(Fe$_{1-x}$Ru$_x$As)$_2$ \cite{Tanabe2} is shown for comparison.
It is apparent that neither Ru nor Mn substitution alters the itinerant characteristics of the Dirac cone above 15\,K in Ba(FeAs)$_2$.
On the other hand, $m^{\ast}$ /$m_0$ for the $x$ = 0.028 and 0.035 samples is greatly enhanced at 2 K compared with the values estimated above 15 K.
The present research clearly shows that the magnetic Mn impurity in Ba(FeAs)$_2$ is the origin of this anomalous enhancement, being in stark contrast with the effect of nonmagnetic impurities in Ba(FeAs)$_2$ \cite{Tanabe2}.
Considering the above discussion, the emergence of the Kondo effect, via band renormalization due to interaction between the localized 3d electrons of Mn and the itinerant Dirac electrons, appears to be a plausible and reasonable interpretation.

\section{Conclusion}
We studied the effect magnetic impurities on the Dirac cone in Ba(Fe$_{1-x}$Mn$_x$As)$_2$ using rapidly quenched single crystals.
Linear magnetoresistance as a consequence of the quantum limit of the Dirac cone, characterized under the SDW transition states, was demonstrated above the crossover magnetic field ($B^{\ast}$).
For $x$ = 0.028 and 0.035, both anomalous enhancement of $B^{\ast}$ and a large decrease in the average mobility ($\mu_{ave}$) were observed at low temperatures.
The inverse of $\mu_{ave}$ (1/$\mu_{ave}$) and the estimated cyclotron effective mass ($m^{\ast}$) showed positive deviations from the conventional dependence of $m^{\ast}$ on the carrier number in the Dirac cones.
A logarithmic increase in electrical resistivity with decreasing temperature, as well as its saturation at further low temperatures, was observed consistent with the the Kondo effect.
We concluded that the Kondo effect occurs between Dirac fermions and Mn 3$d$ localized electrons, leading to an anomalous mass enhancement of the Dirac fermions in Ba(Fe$_{1-x}$Mn$_x$As)$_2$.

\section*{Acknowledgements}
The authors are grateful to Y. Kuramoto, T. Tohyama, S. Takagi, and C. Kim for their useful comments.
This research was partially supported by Grant-in-Aid for Young Scientists (B) (23740251) and GCOE program at Tohoku University.
This work was also supported in part by the approval of the Japan Synchrotron Radiation Research Institute (JASRI).

\begin{figure}[b]
\includegraphics[width=0.6\linewidth]{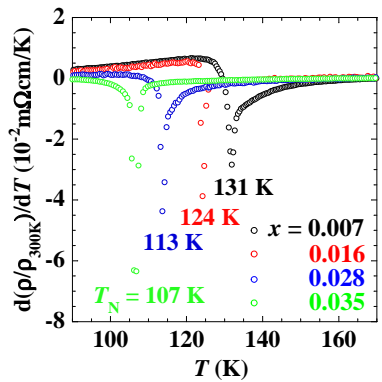}
\caption{(color online) First derivative of normalized resistivity for Ba(Fe$_{1-x}$Mn$_x$As)$_2$.}
\end{figure}

\appendix*
\section{Characterizations of Ba(Fe$_{1-x}$Mn$_x$As)$_2$ single crystals}

\subsection{Powder x-ray diffraction}
In order to evaluate the quality of Ba(Fe$_{1-x}$Mn$_x$As)$_2$ single crystals, synchrotron powder x-ray diffraction (XRD) measurements were performed at SPring-8 beam line BL02B2, using radiation of wave length $\lambda = 0.5 \rm\AA$.
Diffraction patterns were checked for ground single crystals from each batch.
Figure 5 shows refined powder XRD patterns for the Ba(Fe$_{1-x}$Mn$_x$As)$_2$ samples of composition $x$ = 0.007 - 0.035.
Rietveld refinement was performed using General Structure Analysis System (GSAS) with a graphical user interface (EXPGUI) \cite{GSAS,EXPGUI}.
The red curves and pink marker lines indicate the Rietveld fit and the Bragg peak positions of Ba(FeAs)$_2$, respectively.
The blue curves show the difference between the Rietveld fit and the observed intensities.
No detectable impurity phases were found in these compounds, and the diffraction patterns showed good agreement with the results of Rietveld refinement, indicating good quality of our single crystals.
The lattice constants obtained from the Rietveld refinement are shown in table I.
With increasing nominal Mn concentration, both a- and c-axis lattice constants increased, consistent with the previous report \cite{Canfield}.
This indicated that the Mn concentration of the Ba(Fe$_{1-x}$Mn$_x$As)$_2$ single crystals produced continuously increased by increasing the nominal Mn concentration.

\begin{table}[b]
\caption{Sample properties of Ba(Fe$_{1-x}$Mn$_x$As)$_2$.}
\scalebox{1.0}[1.0]{
{\renewcommand\arraystretch{1.0}
\tabcolsep = 1.0mm{
\begin{tabular}{cccccc}
$x$&0&0.007&0.016&0.028&0.035\\
\hline
$x_{\rm nom}$&0&0.005&0.010&0.020&0.025\\
$a({\rm\AA})$&3.9633(0)&3.9759(1)&3.9762(4)&3.9792(1)&3.9805(2)\\
$c({\rm\AA})$&13.022(0)&13.064(1)&13.068(6)&13.079(8)&13.087(3)\\
$T_{\rm N}$&137.2&131.9&124.2&113.7&107.8
\end{tabular}
}}}
\end{table}

\subsection{Temperature dependence of electrical resistivity}
Figure 6 shows the temperature dependence of first derivative of $\rho/\rho_{300\rm K}$, (${\rm d}(\rho/\rho_{300\rm K})/{\rm d}T$).
${\rm d}(\rho/\rho_{300\rm K})/{\rm d}T$ versus T exhibits abrupt anomalies, which can be attributed to the structure/spin-density-wave (SDW) transition for all samples \cite{Canfield2}.
The structure/SDW transition temperatures ($T_{\rm N}$) were extracted based on the anomaly positions, and are displayed in table I.
$T_{\rm N}$ decreases with increasing nominal Mn-concentration $x_{\rm nom}$.
The Mn-concentration $x$ for the present Ba(Fe$_{1-x}$Mn$_x$As)$_2$ single crystals were determined using the $x$ dependence of $T_{\rm N}$ reported by Thaler $et$ $al$. \cite{Canfield} as shown in table I.
It is noted that $\rho/\rho_{300\rm K}$ as shown in Fig. 4 (a) is smaller for $x$ = 0.016 than for $x$ = 0.007, while $\rho/\rho_{300\rm K}$  continuously increases with increasing $x$ in the previous report \cite{Canfield}.
Since the lattice constant, $T_{\rm N}$, magnetic field ($B$) dependence of magnetoresistance ($R_{\rm M}$) and Hall resistivity ($\rho_{yx}$), continuously increase with $x$ in the present experiment, the source of the unusual change observed for $\rho/\rho_{300\rm K}$ at $x$ = 0.016 is unclear.



\end{document}